\documentclass[twocolumn,showpacs,prb]{revtex4-1}
\usepackage{amssymb}
\usepackage{amsmath}
\usepackage{graphicx}
\usepackage{epstopdf}
\usepackage{bm}

\def\fakebold#1{\relax\ifvmode\leavevmode\fi\ifmmode\setbox0=\hbox{$#1$}\else\setbox0=\hbox{#1}\fi\kern-.02em\copy0 \kern-\wd0\kern .04em\copy0 \kern-\wd0\kern-.0125em\raise.02em\box0}

\begin{document}

\title{Bose-Einstein condensates in multiple well potentials from a variational path integral}
\author{Wattana Ratismith$^{\dagger }$}
\author{Holger Hauptmann$^{\ddagger }$}
\author{Walter T. Strunz$^{\ddagger }$}
\affiliation{$^{\dagger }$Energy Research Institute, Chulalongkorn University, Bangkok
10330 Thailand}
\affiliation{$^{\ddagger }$Institut f\"{u}r Theoretische Physik, Technische Universit\"{a}%
t Dresden, 01062 Dresden, Germany}

\begin{abstract}

We apply a path integral variational approach to obtain analytical expressions for condensate wave functions
of an ultracold, interacting trapped Bose gases. As in many recent experiments,
the particles are confined in a 1D or 3D harmonic oscillator trap which is superimposed by a periodic potential in one direction.
Based on the first order cumulant expansion with respect to a harmonic trial action, and emplyoing a mean-field approximation,
optimal variational parameters are obtained by minimizing an analytical expression for the ground state energy.
Our largely analytical results for energy and condensate wave function 
are in good agreement with fully numerical calculations based on the Gross-Pitaevskii equation.

\end{abstract}

\maketitle

\section{Introduction}

%\cite{Morsch,Bronski,Burger,Dae-II}

Ultracold atomic gases trapped in harmonic and periodic potentials are in the focus of current theoretical 
and experimental research  \cite{Jaksch,Lewenstein,Bloch,Morsch}. The number of energetically available potential wells 
can be tuned through the strength of the confining trap, and the parameters of the standing wave light field.

In fact, there are very many investigations for the extreme case of effectively 
just two potential wells \cite{Michael,Shin,Gati,Gati2,Hall,Gati3,Elena,Khan,Milburn,Ananikian,Simon}, 
developing Josephson oscillations for an appropriate choice of parameters. On the other hand, investigations in the
limit of (almost) periodic potentials can rely on Bloch theory. In that case one is often able to chose a description in terms
of the lowest band only \cite{Eiermann, Tristram}.

The influence of atomic interactions in current experiments
can be observed through self-trapping both in double well systems \cite{Michael}, and
in potentials with many potential wells \cite{Tristram, Anker}.
More recently, Trotzky \textit{et al.} \cite{Trotzky} studied relaxation dynamics in an interacting bosonic 
many-body system in an optical lattice combined with
a shallow harmonic trap, such that about fourty potential wells are of relevance for the dynamics. 
Since the number of particles per site is very small, full quantum calculations involving the N-particle state 
can be carried out on the basis of the celebrated Bose-Hubbard model \cite{Jaksch,Lewenstein,Bloch,Morsch}.
 
In this work we are interested in the limit of many particles per well and a mean-field description of the ultracold Bose 
gas in terms of a single condensate wave function $\phi_0({{\bf r}})$ \cite{Dae-Il, Bronski, Burger}. Instead of a direct numerical determination of the 
wave function as the ground state solution of the Gross-Pitaevskii equation, however, we aim at a largely analytical approach with a 
path intergral description as our starting point.

Path integrals have proven indispensable in many areas of physics, ranging
from quantum mechanics and quantum field theory to important applications in both quantum and classical
statistical physics \cite{FeynmanHibbs, Schulman, Kleinert}. In connection to cold atomic gases, a path
integral approach for the quantum field dynamics helps to derive Gross-Pitaevskii-type stochastic equations that
contain damping and thermal fluctuations \cite{Stoof} (see also \cite{Sigi2}). More recently, we investigated the Bose-Einstein condensate
wave function in a double-well potential within the Feynman path integral variational approach \cite{Wattana}.
In the present paper we expand our earlier work to a trapped condensate with an additional periodic potential in one dimension.
Remarkably, our method leads to the same variational energy as obtained from a simple Gaussian ansatz for the wave function.
However, the Feynman path integral variational approach allows to obtain a much improved expression for the condensate wave 
function as shown in this work.

\section{The Feynman Path Integral}

We examine $N$ Bosons with repulsive interaction confined in a three
dimensional harmonic trap, which is superposed with a periodic potential in x direction.
Such a setup is realized in many current experiments \cite{Bloch, Morsch, Anker, Eiermann, Trotzky}.
The N-body potential thus reads
%\cite{FeynmanKleinert, Kleinert,Bachmann, FeynmanHibbs}. 
\begin{eqnarray}\label{nbodypotential}
V &=&\frac{1}{2} \sum\limits_{i=1}^{N}(\Omega _{x}^{2}x_{i}^{2}+\Omega
_{y}^{2}y_{i}^{2}+\Omega _{z}^{2}z_{i}^{2})  \notag \\
&&+\frac{A}{2}\sum\limits_{i=1}^{N}\left( \cos \left( 2kx_{i}\right)
+1\right) +{g}\sum\limits_{i< j}^{N}\delta (\mathbf{r}_{i}-\mathbf{r}_{j}),
\end{eqnarray}%
where $\Omega _{x},\Omega _{y},$ and $\Omega _{z}$ are harmonic oscillator frequencies of the trap, 
$A$ is proportional to the intensity of a standing wave laser beam and $k$ the corresponding wave number.
The interatomic interaction is taken care of by an effective delta-shaped pseudopotential where 
$g$
% =4 \pi a \hbar^2/ m$ 
is proportional to the s-wave scattering length $a>0$.
We exploit a harmonic trial Lagrangian
\begin{equation}
L_{0}=\frac{1}{2}\sum\limits_{i=1}^{N}(\overset{\cdot }{x}_{i}^{2}+\overset{%
\cdot }{y}_{i}^{2}+\overset{\cdot }{z}_{i}^{2})-\frac{1}{2}%
\sum\limits_{i=1}^{N}(\omega _{x}^{2}x_{i}^{2}+\omega
_{y}^{2}y_{i}^{2}+\omega _{z}^{2}z_{i}^{2}),
\end{equation}
for which the propagator is known in closed form, and as a path integral may be written as
\begin{align}
P_{0}&=\int_{x_i,0}^{x_f,t}D^N[x(\tau)] \exp \left( iS_{0}\right) %\\
%&= \prod_{i=x,y,z} \limits \sqrt{\frac{m \omega_i}{2 \pi i \hbar \sin(\omega_i t)}} \exp(\frac{i m \omega_i}{2 \hbar \sin(\omega t)}(r_i^2+r_i^2)\cos(\omega_i t)-2 r_i r_i )
\end{align}%
with $S_0=\int d \tau L_0$.
Later, $\omega _{x},\omega _{y},\omega _{z}$ are considered to be variational parameters. 
In the following, we calculate the propagator for the $N$-body case with potential $V$ from
(\ref{nbodypotential}). Since
\begin{align}
 \nonumber
 P&= \langle x_f | \exp(-i H t) | x_i \rangle \\
 &=\sum_{n=0}^\infty \limits \exp(-i E_n t) \phi_n(x_f) \phi^*_n(x_i), 
\end{align}
and replacing $ t \rightarrow -i\tau $, we find
the ground state energy and the condensate wave function in the limit $\tau\rightarrow \infty$.
With the harmonic oscillator as a trial system, we can express the propagator of the entire system as 
\begin{align}
\nonumber
P &=\int_{x\left( 0\right) }^{x\left( t\right) }D^N[x(\tau)] \exp \left( iS \right) \\
&=: P_0\left\langle \exp \left[ i\left( S-S_{0}\right) \right] \right\rangle_{S_{0}}.  \label{p1} 
\end{align}%
The bracket $\left\langle \ldots \right\rangle _{S_{0}}=\int D^N[x] \exp[iS_0] (\ldots)/P_0$ denotes 
the ``average'' with respect to the trial
``probability'' density functional $\exp(i S_0)$.
Similar to related expansions in probability theory, and following
Bachmann et al. \cite{Bachmann}, this propagator can be expanded in terms of cumulants.
In first order we get from equ. (\ref{p1})
\begin{equation}
P \approx P_{0}\exp \left[ i\left\langle S-S_{0}\right\rangle _{S_{0}}\right] .
\label{p2}
\end{equation}%

Similar to standard
probability theory it is possible to obtain mean values from
generating functionals \cite{FeynmanHibbs} 
\begin{align}
 \Phi[f(t)]=\left\langle \exp \left[ i\int_0^t \limits f(\tau )x(\tau)d\tau \right] \right\rangle _{S_{0}} .
\end{align}
Average values of $x(\tau)$, for instance are obtained from 
\begin{equation}
\left\langle x(\tau )\right\rangle _{S_{0}}=-i\left. \frac{\delta  \Phi[f(\tau)]  }{\delta
f(\tau )}\right\vert _{f=0}\; \label{generating}
\end{equation}
and
\begin{align}
 \left\langle x(\tau )^2\right\rangle _{S_{0}}=(-i)^2 \left. \frac{\delta^2 \Phi[f(\tau)] }{\delta
f(\tau )^2}\right\vert _{f=0}\;.  
\end{align}
The generating functional can be expressed in terms of the difference between the action of a driven harmonic oscillator $S_0'$ and 
the action of a usual harmonic oscillator $S_0$
\begin{align}
 \Phi[f(t)]=\left\langle \exp \left[ i(S_0'-S_0) \right] \right\rangle _{S_{0}}= \exp \left[ i(S_{cl}'-S_{cl}) \right] .
\end{align}
Here $S_{0}^{\prime }=S_{0}+\int f(\tau )x(\tau )d\tau $ represents the action of a driven oscillator with external force $f(\tau)$.
For harmonic systems the path integral can be evaluated in closed form. One finds
$P \sim \exp{(iS_{cl})}$ for the propagator with the action $S_{cl}=S_0[x_{cl}]$ along the classical path.
Thus, 
\begin{equation}
\left\langle x(\tau )\right\rangle _{S_{0}}= \left.\frac{\delta S_{cl}^{\prime }}{\delta f(\tau )} \right\vert _{f=0}
\end{equation}
and
\begin{align}
 \left\langle x (\tau )^{2}\right\rangle_{S_0} =\left. \left[ -i\frac{\delta ^{2}S_{cl}^{\prime }}{\delta f(\tau )^{2}}+\left( \frac{\delta S_{cl}^{\prime }}{\delta f(\tau )}\right) ^{2}\right] \right\vert _{f=0}.
\end{align}
The analytical expression in one dimension, for simplicity, reads \cite{FeynmanHibbs}
\begin{align} \nonumber
 &S_{cl}^{\prime }=\frac{\omega_x}{2 \sin(\omega_x t)} \Bigg( \left((x_i^2+x_f^2)\cos(\omega t)-2 x_i x_f \right) \Bigg. \\ \nonumber
 & + \frac{2 x_f}{\omega_x} \int_0^t \limits d \tau  f(\tau) \sin \omega_x \tau +\frac{2 x_i}{\omega_x} \int_0^t \limits d \tau  f(\tau) \sin \omega_x (t-\tau) \\
 &\Bigg. -\frac{2}{ \omega_x^2} \int_0^t \limits d\tau \int_0^\tau \limits d s \; f(s) f(\tau) \sin \omega_x \tau \sin \omega_x (t-\tau) \Bigg)
\end{align}
and leads us directly to
the Green function
\begin{align}
 g(t, \tau)=\langle x^2 \rangle_{S_0} -\langle x \rangle_{S_0}^2 =\frac{i}{\omega_x} \frac{\sin \omega_x (t-\tau) \sin \omega_x \tau }{\sin \omega_x t }.
\end{align}
Moreover, we have to calculate averages of more complicated functions of $x(\tau )$.
These can be obtained along similar lines from the generating functional.
The contribution of the cosine potential, for instance, is
\begin{align}
\left\langle \cos \left( 2 k x\right) \right\rangle _{S_{0}} =\exp\left[-2 k^{2} g(t,\tau) \right] \cos \left( 2k\left\langle x\right\rangle _{S_{0}}\right).
\end{align}
For the interatomic interaction, we use the Fourier representation
\begin{align}
\delta(x_i-x_j) = \frac{1}{2\pi}\int dq\,\exp{[iq(x_i-x_j)]}.
\end{align}
Taking the Gaussian average, we find
\begin{align}
 & \left\langle \exp \left[ iq\left( {x}_{i}-{x}_{j}\right) \right]
 \right\rangle _{S_{0}}  \notag \\
 = & \exp \left[ -q^{2}\left( g\left( t ,\tau \right) +\left\langle
 x_{i}\right\rangle _{S_{0}}\left\langle x_{j}\right\rangle
 _{S_{0}}-\left\langle x_{i}x_{j}\right\rangle _{S_{0}}\right) \right] . 
\end{align}
Using the mean field factorization $\left\langle x_{i}x_{j}\right\rangle
_{S_{0}}=\left\langle x_{i}\right\rangle _{S_{0}}\left\langle
x_{j}\right\rangle _{S_{0}}$, we obtain 
\begin{equation}
\left\langle \delta ({x}_{i}(\tau )-{x}_{j}(\tau ))\right\rangle _{S_{0}}=%
\sqrt{\frac{1}{4\pi g\left( t,\tau \right) }}.
\end{equation}%

Ground state properties (ground state energy and condensate wave
function) can be obtained by taking the limit $t\rightarrow -i\,\infty $ of
the full expression of the propagator in Eq. (\ref{p2}). In our case, the
result for the energy is 
\begin{align}
\nonumber
E_{0}\left( \omega _{x},\omega _{y},\omega _{z}\right) &/N = \frac{\omega _{x}}{4}+\frac{\omega _{y}}{4}+\frac{\omega _{z}}{4}+\frac{\Omega _{x}^{2}}{4\omega _{x}}+\frac{\Omega _{y}^{2}}{4\omega _{y}} \\
\nonumber &+\frac{\Omega _{z}^{2}}{4\omega _{z}} +\frac{A}{2}+\frac{A}{2}\exp \left( -\frac{k^{2}}{\omega _{x}}\right) \\
&+\frac{g\left( N-1\right) }{2}\left( \frac{1}{2\pi }\right) ^{3/2}\sqrt{\omega _{x}\omega _{y}\omega _{z}}.  \label{E3D}
\end{align}
Remarkably, this expression can also be obtained from a more direct variational approach with Gaussian trail wave functions 
with variances $1/2w_x$, $1/2w_y$ and $1/2w_z$ for the respective directions. 

Our variational condensate wave function
can be read off in the limit  $t\rightarrow -i\,\infty $ and we obtain
\begin{align}
\nonumber
\phi _{0}\left( x,y,z\right) \sim \exp \left[ -\left( \frac{\Omega _{x}^{2}}{4\omega _{x}}+\frac{\omega _{x}}{4}\right)x^{2}  \right. \ \ \ \ \ \ \ \ \ \ \ \ \\
\left. -\left( \frac{\Omega _{y}^{2}}{4\omega _{y}}+\frac{\omega _{y}}{4}\right) y^{2} -\left( \frac{\Omega _{z}^{2}}{4\omega _{z}}+\frac{\omega _{z}}{4}\right) z^2 
-\frac{A}{2}\lambda _{0}\left( x\right)\right]   \label{coswave}
\end{align}
with
\begin{align}
\nonumber
\lambda _{0}\left( x\right) =&-\frac{1}{\omega _{x}} \left( \gamma -\text{Ci}\left( 2kx\right) +\ln(2kx) \right) \\
&+\frac{1}{2}\sum\limits_{j=1}^{\infty }\frac{2^{2j+1}\omega _{x}^{j-1}e^{-\frac{k^{2}}{\omega _{x}}}}{(2j)^{2}}x^{2j}\theta \left( j\right) ,
\end{align}
\begin{equation}
\theta \left( j\right) =\frac{\Gamma \left( j+1\right) -\Gamma
\left( j+1,-\frac{k^{2}}{\omega _{x}}\right) }{\Gamma \left(2j\right) }.
\end{equation}
Here $\gamma \simeq 0.577216 $ is Euler's constant, Ci$\left( x\right)$ is the cosine integral function, $\Gamma \left( x\right)$ is the Euler gamma function and $\Gamma \left( x,y \right)$ is the incomplete Euler gamma function.
We can normalize this wave function by using the condition $\int \left\vert \phi _{0}\left( \mathbf{r}\right) \right\vert ^{2}d^{3}\mathbf{r}=1.$
The condensate wave function decomposes in a product of functions for each dimension.
In $y$- and $z$-dimension it turns out to be Gaussian, as could be expected because the condensate is harmonically trapped in these directions.
However, in $x$-direction due to the additional periodic potential, a contribution $A \lambda_0(x)/2$ in the exponential 
modifies the Gaussian shape significantly.
A damped periodic modulation of the condensate wave function along $x$ is obtained.
We numerically minimize the energy functional with respect to the variational parameters and obtain the optimal $\omega_x$, $\omega_y$ and $\omega_z$,
which are then used to compute the condensate wave function $\phi_0({\bf r})$.

\section{Comparison with full numerics}

Next we compare the quality of our results from the Feynman path integral variational approach with a numerical
mean-field calculation obtained through imaginary-time propagation of the Gross-Pitaevskii equation.
The comparision is done with 1D and 3D isotropic, and 3D anisotropic harmonic traps.
\begin{figure}[tbp]
\begin{center}
%\hspace{-0.49cm}
\includegraphics[scale=0.8] {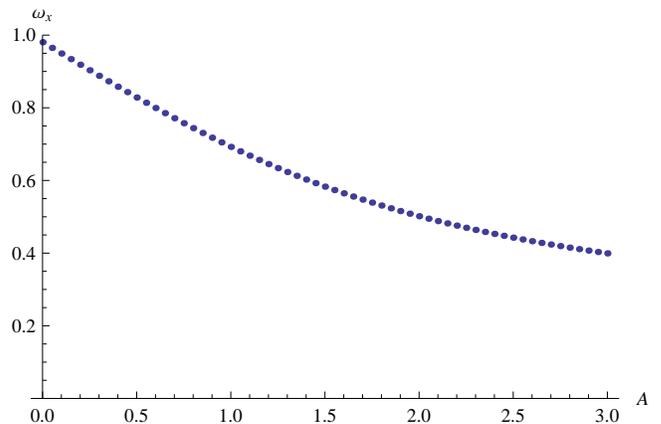}
\end{center}
\caption{Minimizing parameter $\protect\omega_x $ for the 1D system as a function of $A$ for $k=0.744,\Omega _{x}=1,$ $gN=0.1$.}
\end{figure}
Using the Feynman path integral approach in one dimension, we get the ground state energy of the entire system 
\begin{eqnarray}
\frac{E_{0}\left( \omega _{x}\right) }{N} &=&\frac{\omega _{x}}{4}+\frac{%
\Omega _{x}^{2}}{4\omega _{x}}+\frac{A}{2}+\frac{A}{2}\exp \left( -\frac{%
k^{2}}{\omega _{x}}\right)  \notag \\
&&+\frac{g\left( N-1\right) }{2}\sqrt{\frac{\omega _{x}}{2\pi }}.
\end{eqnarray}%
The corresponding condensate wave function is%
\begin{equation}
\phi _{0}\left( x\right) \sim \exp \left[ -\left( \frac{\Omega _{x}^{2}}{%
4\omega _{x}}+\frac{\omega _{x}}{4}\right) x^{2}+\frac{A}{2}\lambda
_{0}\left( x\right) \right] .  \label{wave2}
\end{equation}%
We minimize this energy by solving \ $\partial E_{0}\left(\omega _{x}\right) /\partial \omega _{x}=0$ for a small repulsive self interaction $g(N-1) \approx gN=0.1$.
The relation between the minimizing parameter $\omega _{x}$ and $A$ is shown in Fig. 1.
The comparison between variational energy and numerically exact Gross-Pitaevski energy is displayed in  Fig. 2.
\begin{figure}[tbp]
\begin{center}
%\hspace{-0.49cm}
\includegraphics[scale=0.8] {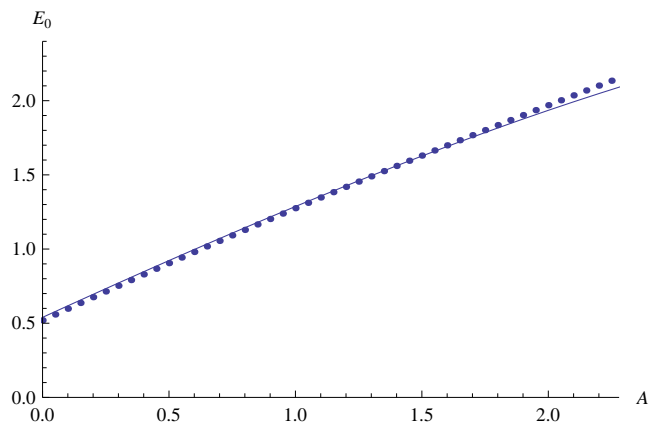}
\end{center}
\caption{Ground state energy as a function of $A$ for the 1D system. The solid line is the numerical solution of the Gross-Pitaevskii equation and the dots are obtained from path
integral theory for $k=0.744,\Omega _{x}=1,$ $gN=0.1$.}
\end{figure}
We next insert these optimal values for $\omega _{x}$ into Eq. (\ref{wave2}) and normalize. In Fig. 3 we compare this variational wave function (\ref{wave2}) with
the numerically exact solution of the Gross-Pitaevskii equation for various parameters. The corresponding potentials range from almost double-well type
(for strong trapping confinement, $\Omega = 1$) to genuine multiple-well shape for weaker traps ($\Omega = 0.1, 0.02$). In the first case,
the condensate wave function Eq. (\ref{wave2}) is separated into two peaks which are symmetrically centered around the origin. In the latter cases,
the condensate wave function reflects the multiple-well structure of the confining potential (see Fig. 3). 
\begin{figure}[tbp]
\begin{center}
%\hspace{-0.49cm}
\includegraphics[scale=0.5] {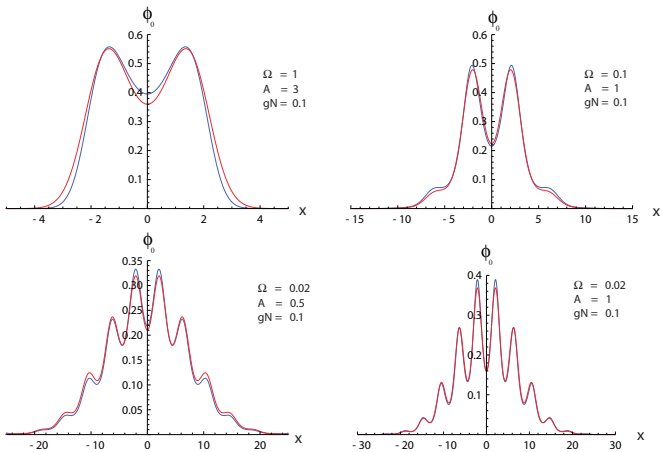}
\end{center}
\caption{Condensate wave function  $\phi_0(x)$ for the one dimensional system. The red lines are the results of imaginary time propagation of the Gross-Pitaevskii 
equation, the blue lines are calculated by our variational path integral approach. }
\end{figure}

We observe that over a wide range of potential parameters, 
the condensate wave function from the path integral variational approach agrees very well with the numerical solution of the Gross-Pitaevskii equation. 

Equivalently, Bose-Einstein condensate wave functions in 3D can be determined by minimizing the ground state energy in 
equ. (\ref{E3D}), see Fig. 4 for the choices $\Omega=\Omega_x =\Omega_y =\Omega_z = 1$ and $gN=1$.
We obtain the values of the minimizing parameters and from these the value of the ground state energy.
Inserting these optimal $\omega_{x},\omega _{y},\omega _{z}$ into equ. (\ref{coswave}) and normalizing the wave function, 
we obtain $\phi _{0}$.
Cuts of the densities $n_0=\phi _{0}^{2}$ at $z=0$ are shown in Fig. 5 at $gN=1$ for various trap parameters: top: $\Omega=1, A=3$, double-well type;
middle and bottom: $\Omega=0.02$ multiple-well type, with $A=0.5$ (middle) and $A=1.0$ (bottom).
Again we observe remarkable agreement with numerically exact solutions of 
the Gross-Pitaevskii equation. 
\begin{figure}[tbp]
\begin{center}
%\hspace{-0.49cm}
\includegraphics[scale=0.8] {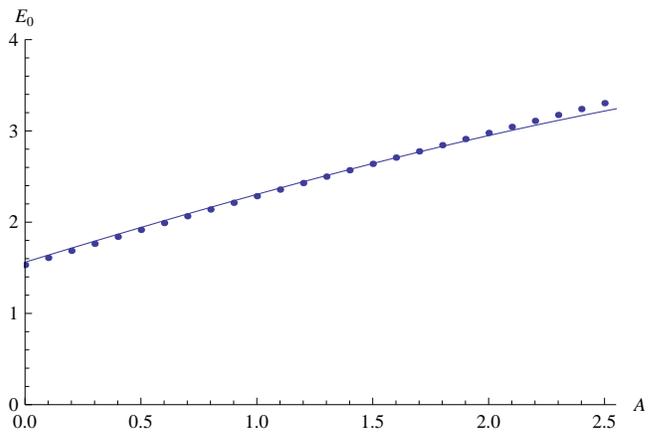}
\end{center}
\caption{Ground state energy in 3D as a function of $A$. The
solid line is calculated from the Gross-Pitaevskii equation and the dots are
calculated by path integral variational theory for $k=0.744,\Omega _{x}=\Omega _{y}=\Omega
_{z}=1,$ $gN=1$.}
\end{figure}
\begin{figure}[tbp]
\begin{center}
%\hspace{-0.49cm}
\includegraphics[scale=0.25] {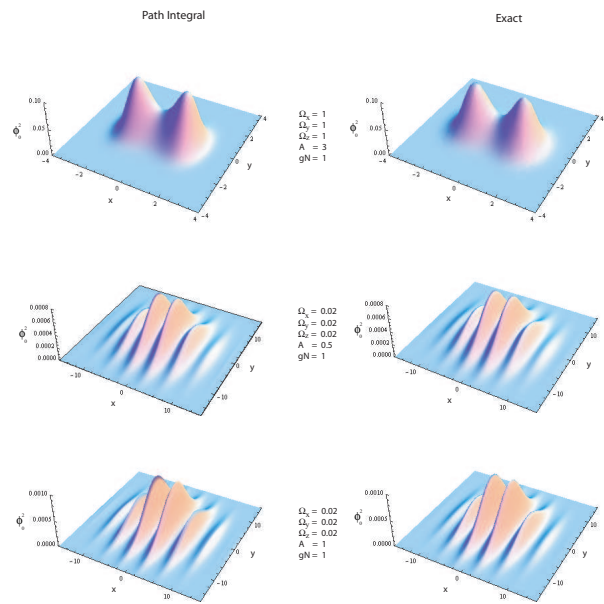}
\end{center}
\caption{Density cuts of the variational condensate wave functions at $z=0$ in 3D (left column) 
compared with numerically exact solutions
of the Gross-Pitaevskii equation for an isotropic confining trap (right column).}
\end{figure}
Finally, in Fig. 6 we exhibit results for anisotropic confining traps (interaction parameter $gN=1$).
The first graphs correspond to a pan cake type condensate in $x$-direction ($\Omega_{x}=1,\Omega _{y}=\Omega _{z}=0.05$) 
with a fairly strong cosine potential $A=3$, which leads to a 
very narrow and double-well-like confinement in $x$ direction (Fig. 6, top).
As a second example (Fig. 6, middle) we present plots for a cigar type confinement in $x$-direction, 
($\Omega_{x}=0.02,\Omega _{y}=\Omega _{z}=1$) and barrier hight $A=1$.
Thirdly, there is a slightly anisotropic and weakly confining example with $\Omega_{x}=0.02,\Omega _{y}=\Omega _{z}=0.01$.
In this case the barrier hight is $A=1$ so that many potential wells are populated.
As before, the ground state energies of these three examples and the minimizing parameters can be determined from equ. (\ref{coswave}) 
with results shown in Table 1.

\begin{table}[tbp]
\begin{center}
\begin{tabular}{|l|c|c|c|c|c|c|c|c|}
\hline
$A$ & $\Omega _{x}$ & $\Omega _{y}$ & $\Omega _{z}$ & $\omega _{x}$ & $%
\omega _{y}$ & $\omega _{z}$ & $E_{0}$ & $E_{0,GP} $
\\ \hline
3 & 1 & 0.05 & 0.05 & 0.4019 & 0.0490 & 0.0490 & 2.652 & 2.463 \\ \hline
1 & 0.02 & 1 & 1 & 0.01640 & 0.9960 & 0.9960 & 1.514 & 1.415 \\ \hline
1 & 0.02 & 0.01 & 0.01 & 0.01995 & 0.00996 & 0.00996 & 0.520 & 0.414 \\ 
\hline
\end{tabular}%
\end{center}
\caption{Optimizing variational parameters $\omega_x$, $\omega_y$ and $\omega_z$ for the three examples for 3D anisotropic traps.
The corresponding ground state energies from path integral theory $E_0$ and from Gross-Pitaevskii numerics $E_{0,GP}$ are in fair 
agreement. }
\end{table}
\begin{figure}[tbp]
\begin{center}
%\hspace{-0.49cm}
\includegraphics[scale=0.25] {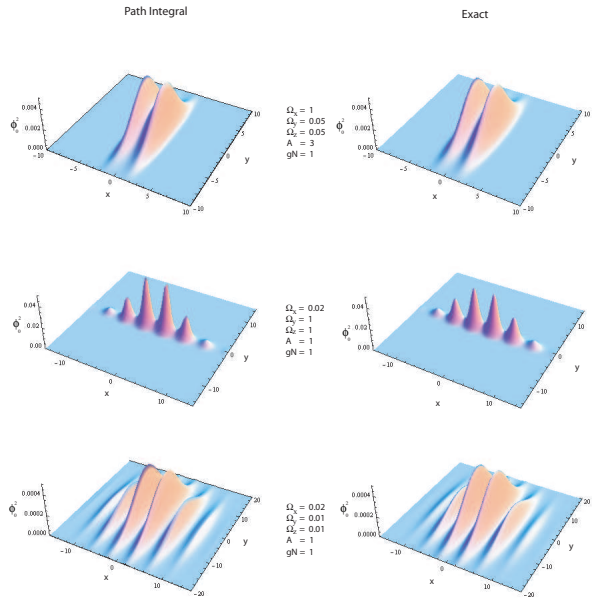}
\end{center}
\caption{Density cuts of the condensates at $z=0$ in 3D compared with the exact solutions
of the Gross-Pitaevskii equation for anisotropic traps.}
\end{figure}

\section{Conclusion}

We have studied Bose-Einstein condensates in a multiple-well potential applying Feynman path integral 
variational theory. As in recent experiments, the trap consists of a harmonic confinement superimposed by a periodic (cosine)
standing wave field.
The calculations are carried out within the first cumulant approximation measured with respect to a harmonic trial action.
The advantage of this method is that we obtain analytical expressions for ground state energy and condensate wave function.
These results are in remarkable
agreement with fully numerical calculations based on the Gross-Pitaevskii equation. We concentrate on a fairly weak interaction
parameter $gN$, no larger than unitiy. For stronger interactions deviations from Gross-Pitaevskii theory begin to appear. It is very remarkable that
our Feynman path integral variational approach leads to the same ground state energy as a purely Gaussian trail wave function.
However, the path integral approach offers a highly non-trivial expression for the condensate wave function with strongly non-Gaussian
shape. In fact, we find very good agreement with the exact condensate wave function as obtained from a numerical solution
of the Gross-Pitaevskii equation.

In future investigations we aim to drop the factorization approximation $\langle x_i x_j\rangle \approx \langle x_i\rangle\langle x_j\rangle$,
such as to be able to go beyond mean-field Gross-Pitaevskii theory.

\begin{acknowledgments}
The authors thank the Erasmus Mundus Action 2, the German Academic Exchange Service
(DAAD) and the International Max Planck Research School (IMPRS) at the MPI for the Physics of Complex Systems (Dresden) 
for their support. Fruitful discussions with Holger Kantz, Virulh Sa-yakanit, John S. Briggs and Sigi Heller are gratefully acknowleged.\\
\\
\\
\end{acknowledgments}

\appendix

\end{document}